\documentclass[epj]{svjour}
\usepackage{graphics}
\newcommand{\comm}[1]{}

\newcommand{\kl}{{\rm K_{L}}}

\newcommand{\ks}{{\rm K_{S}}}

\newcommand{\kzb}{\mbox{$\rm \bar{K}^{0}$}~}
\newcommand{\kz}{\mbox{$\rm K^{0}$}~}

\newcommand{\threepio} {\mathrm{3 \pi ^0}}

\newcommand{\KLPP} {\kl \rightarrow \pi \pi}
\newcommand{\KLPPP} {\kl \rightarrow \threepio}

\newcommand{\KSPPP} {\ks \rightarrow \threepio}
\newcommand{\KPPP} {K \rightarrow \threepio}
\newcommand{\ETAooo} {\eta_{ooo}}
\newcommand{\PreserveBackslash}[1]{\let\temp=\\#1\let\\=\temp}

\newcommand{\definmath}[2] {\def#1{\ifmmode#2\else\ensuremath{\mathrm{#2}} \fi}}

\begin{document}
\title{Search for CP-Violation in $\KSPPP$ decays with the NA48 detector}
\author{Nicol\`o Cartiglia}
%
%
\institute{ INFN, Turin, Italy}
\date{Received: date / Revised version: date}
%
\abstract{ The decay $\KSPPP$ is forbidden by CP conservation. Using a  sample of more than 6 million $\KPPP$ decays, the NA48 Collaboration has improved the  limit on $\eta_{000} = A(\KSPPP)/A(\KLPPP)$ and on the branching ratio $Br(\KSPPP) $ by about one order of magnitude. Using this  result and the Bell-Steinberger relation, a new limit on the equality of the $\kz$ and $\kzb$ masses is obtained improving by about 40\% the test of  CPT conservation in the mixing of neutral kaons.
\PACS{
      {PACS-key}{discribing text of that key}   \and
      {PACS-key}{discribing text of that key}
     } 
} 
\maketitle
\section{Introduction}
\label{intro}
The NA48 experiment was optimized to measure the value of $Re(\epsilon\prime/\epsilon)$, i.e. the ratio of {\it direct} over {\it indirect} CP violation in the kaon sector ~\cite{epsilon},~\cite{epsilon1}. It took data between 1997 and 2001. The data taken during this period have been also used to perform a variety of other measurements of CP violation ($\ETAooo$ and $K_{e3}$ charge asymmetry), mass and lifetime ($K$ and $\eta$ mass, $K$ lifetime) and kaon and hyperon rare decays~\cite{arci}.
\section{The NA48 Detector}
\label{detector}
\begin{figure}
\resizebox{0.5\textwidth}{!}{%
\includegraphics{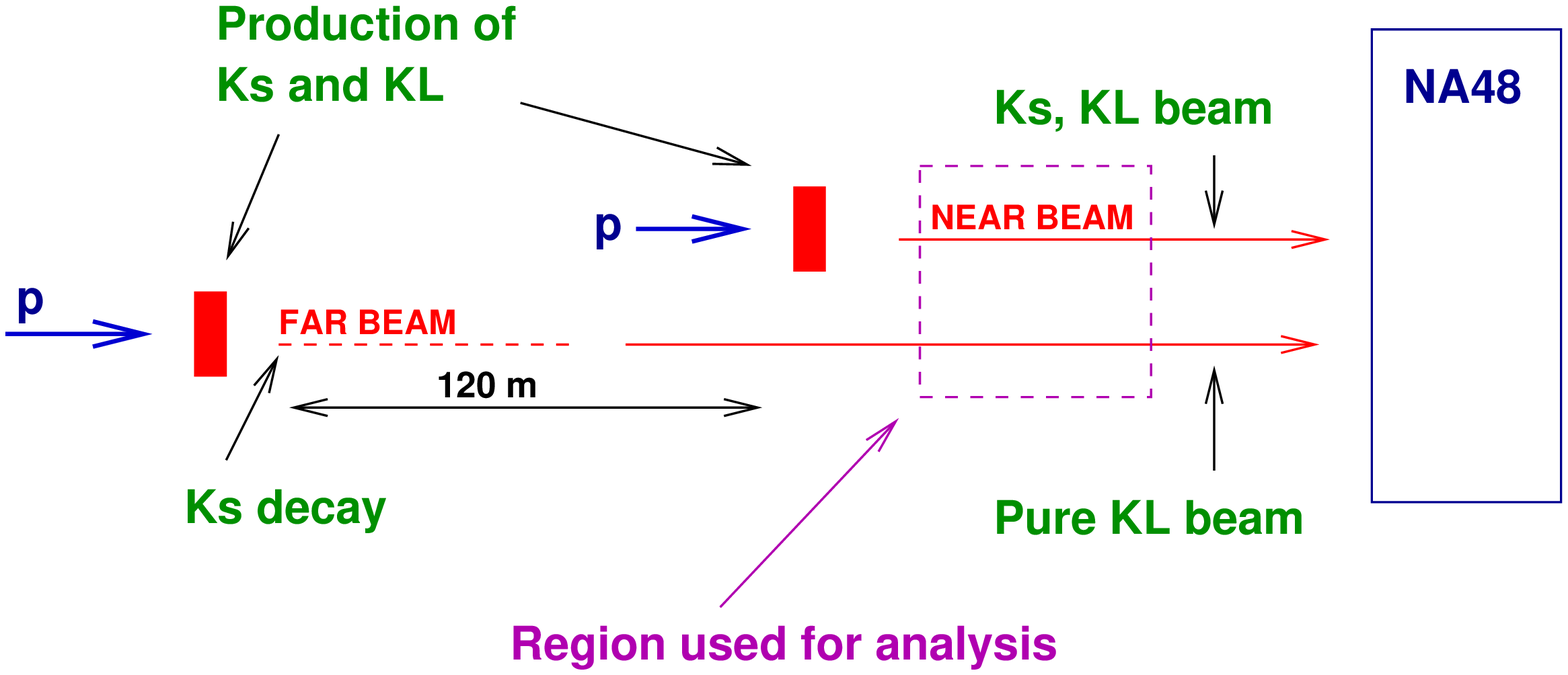}
}
\caption{The beam structure of the NA48 experiment}
\label{fig:beams}       
\end{figure}

The NA48 experiment is a fixed target experiment which uses two concurrent and quasi overlapping beams of kaons, Fig.~\ref{fig:beams}. One kaon beam  (called FAR beam) is produced 126~m upstream the other beam and by the time it reaches the decay region all its $\ks$ mesons have decayed away. The second  kaon beam (called NEAR beam) is produced only 6~m before the decay region and therefore  contains both  $\ks$ and  $\kl$. The kaons are produced by a primary 450~GeV (400 GeV in 2001) proton beam ($\sim 1.5 \cdot 10^{12} $ per spill on the FAR target and $\sim 3. \cdot 10^{7} $ on the NEAR target ) impinging on a 400 mm long, 2 mm diameter rod of beryllium. Charged particles from decays are measured by a magnetic spectrometer composed by four drift chambers with a dipole magnet between the second and third one which introduces a momentum kick of 265 MeV/c in the horizontal plane. The space point resolution is $\sim~95\mu m$ and the momentum resolution is $\sigma(p)/p = 0.48\% \oplus 0.009\% * p [GeV]$ (2001 values). The spectrometer is followed by a liquid kripton calorimeter 27 radiation length long with an energy resolution of $\sigma(E)/E = (3.2 \pm 0.2)\%/\sqrt{E} \oplus (9\pm 1)\%/E \oplus (0.42 \pm 0.05) \%$. The detector is complemented by an hadronic calorimeter, a muon detector, fast hodoscopes for triggering, a proton tagging system, beam monitors. A full description can be found in \cite{epsilon},\cite{epsilon1}.

\section{The Kaon system}
\label{Kaon}

The $\kz, \kzb$ flavour eigenstates are created by strong interaction. These states  mix and propagate as mass eigenstates, $\ks$ and $\kl$, which are a superposition of CP eigenstates: $\ks $ is a quasi pure $CP = 1$ state and $\kl$ a quasi pure $CP = -1$ state, Tab.~\ref{tab:ka}. There is therefore a mismatch between the CP and the mass eigenstates which allows  both $\ks$ and $\kl$ to decay into states of opposite CP.

\begin{figure}[t]
\resizebox{0.45\textwidth}{!}{%
\includegraphics{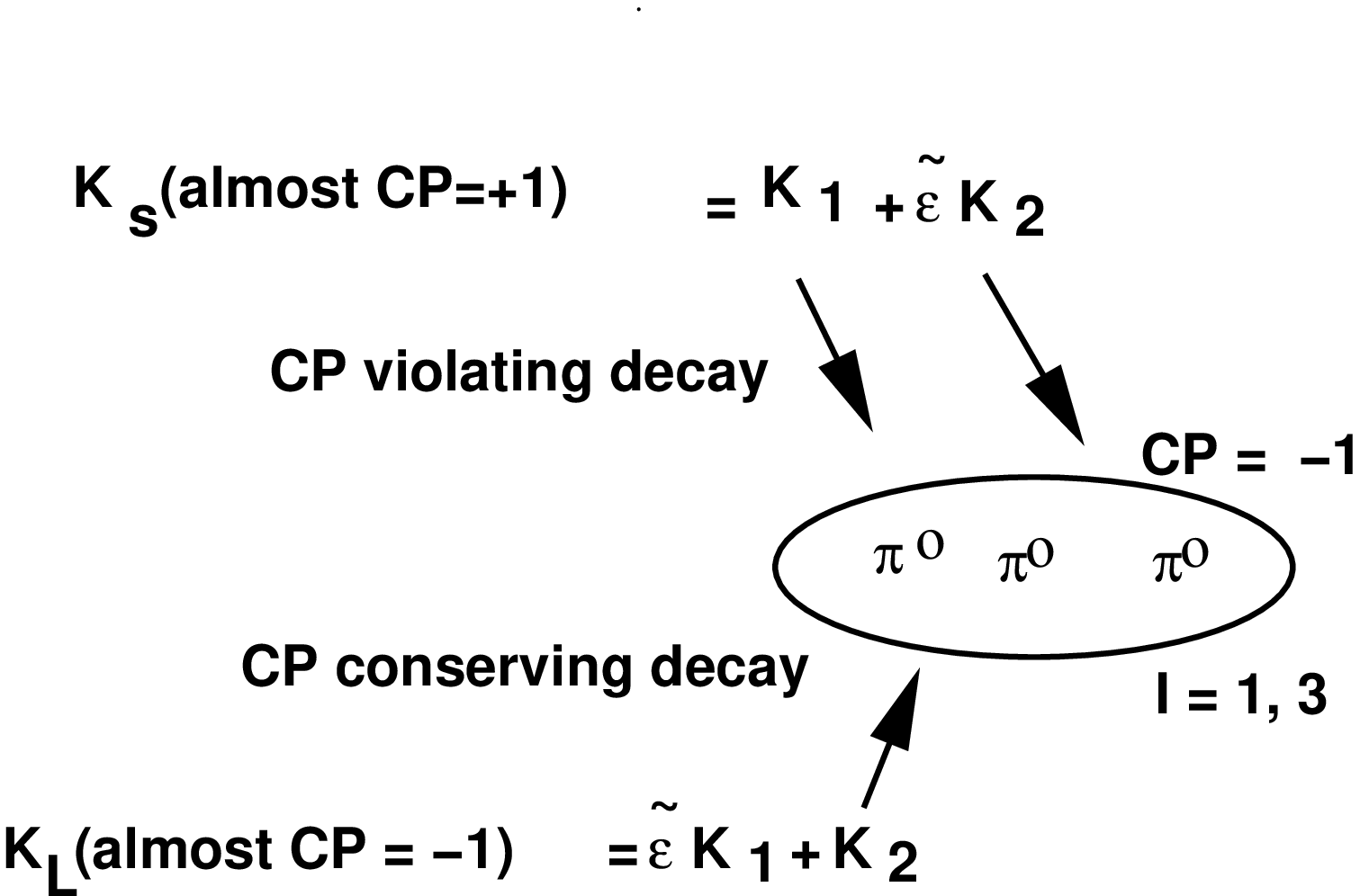}
}
\caption{$\KSPPP$ and $\KLPPP$ decay mode}
\label{fig:cp2}       
\end{figure}

Consider now the decay $\KPPP$. Let's calculate the P, C and I values of  a $|\threepio>$ state. The parity is given by  $P|3\pi^o> = (-1)^l(-1)^L (-1)^3|3\pi^o>$ where $l$ is the angular momentum of a pair of $\pi^o$, $L$ is the angular momentum of the third  $\pi^o$ with respect of this pair and $(-1)^3$ is the intrinsic parity of a $|\threepio>$ state. Since the total angular momentum  is $J=0$  then $l = L$ and  $P|3\pi^o> = (-1)^{2l} (-1)^3|3\pi^o> = - |3\pi^o>$. The charge conjugation operation on a $\pi^o$ does not change its state, $C|\pi^o> = |\pi^o>$ so we have $C|3\pi^o> = (+1)^3|3\pi^o> = + |3\pi^o>$.  The isospin values of  a  $|\threepio>$ state are I=1 and I=3, which are both  symmetric. The  total  wavefunction $|\threepio> = |spin> |space> |isospin> $ must be symmetric (three identical bosons) so both isospin values are allowed (the $|spin>|space>$ component, with $S=0$ and $l+L = 0$ is of course symmetric). We have then:  $CP|3\pi^o> = - |3\pi^o>$ with $\KLPPP$  a CP conserving decay and  $\KSPPP$  a CP violating decay, both with $\Delta I = 1/2, 5/2$, Fig.~\ref{fig:cp2}. 

\begin{table}
\caption{Kaon Eigenstates}
\label{tab:ka}
\begin{tabular}{lcc} 
Eigenstate & expression    & CP value \\ 
\noalign{\smallskip}\hline\noalign{\smallskip}
 Strong & $\kzb(\bar{d}s),~\kz(d\bar{s})$ &  \\
 CP   & $K_1 \propto (\kz + \kzb)$ & +1 \\
 CP   & $K_2 \propto (\kz - \kzb)$ & -1 \\
 Mass & $\ks \propto K_1+\epsilon K_2$ & Almost +1  \\
 Mass & $\kl \propto \epsilon K_1 + K_2$ & Almost -1  \\
\noalign{\smallskip}\hline
\end{tabular}
\end{table}

\section{$\ETAooo$}
\label{eta}
In order to quantify the strength of CP violation in the  $\KSPPP$ decay the following quantity has been introduced~\cite{pdg}: 
\begin{equation}\ETAooo = \frac{A(\KSPPP)}{A(\KLPPP)}.\end{equation}
Assuming CPT invariance, using the  Wu-Yang phase convention ($Im (a_0) = 0 \to \epsilon = \tilde{\epsilon}$) and  ignoring transition into I=3 final states $\ETAooo$ can be rewritten as:
\begin{equation}\ETAooo =  \epsilon + i \frac{Im(a_1)}{Re(a_1)}\label{eq:eta}\end{equation}
where $a_1$ is the weak  amplitude for $\kz$ to decay into I=1 final states and $\epsilon$ can be derived from the $\KLPP$ decay. 
In eq.~\ref{eq:eta} $Re(\ETAooo) = Re(\epsilon)$ so it's only the immaginary part which is sensitive to direct CP violation.

\section{The method}
\label{method}
\begin{figure}[h]
\resizebox{0.5\textwidth}{!}{%
\includegraphics{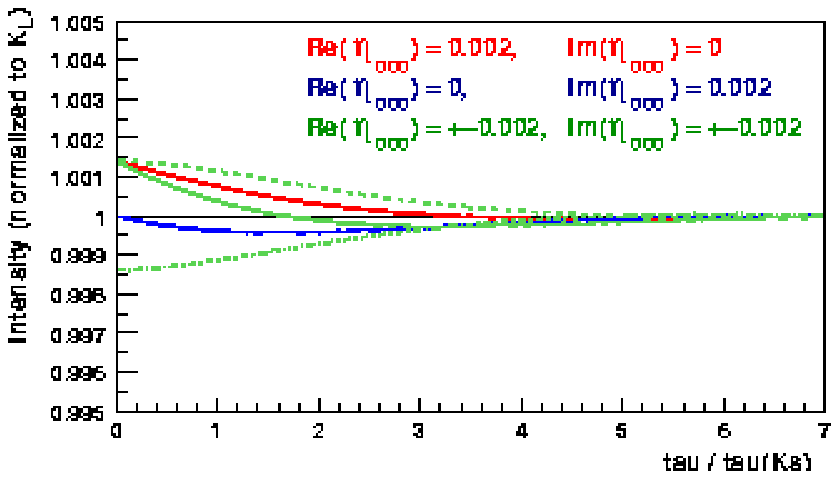}
}
\caption{Interference pattern for different values of  $\ETAooo$ normalized to $\KLPPP$.}
\label{fig:int}       
\end{figure}

Given the very small (still unknown) branching fraction it's very hard to measure directly the decay $\KSPPP$. However, it's  possible to see it's presence since it interferes with the much larger decay  $\KLPPP$: given a $\ks+\kl$ beam, the intensity of $3\pi^o$ decay is given by
$$ I_{3\pi^o} (t) \propto  \underbrace{e^{-\Gamma_Lt }}+\underbrace{|\eta_{ooo}|^2 e^{-\Gamma_St}+}$$ 
\hspace{3.cm} $K_L$  decay \hspace{0.5cm} $K_S$  decay
$$ \underbrace{+2D(p)[Re(\ETAooo)cos\Delta m t - Im(\ETAooo)sin\Delta m t] e^{0.5 (\Gamma_S + \Gamma_L)t}}$$   \vspace{-0.1cm} 
\hspace{2cm} interference $K_S - K_L$ \\
\vspace{0.1cm} \\
 where $D(p) = N(\kz - \kzb)/N(\kz + \kzb) \sim 0.35 $, the dilution factor, parametrizes the $\kz, \kzb$ production asymmetry as a function of the kaon momentum. The maximum interference is at the target and most of the effect is contained within the first 2 $\ks$ lifetime. The interference pattern is superposed over a large $\KLPPP$ signal and it can be positive or negative depending on the value of $\ETAooo$, Fig.~\ref{fig:int}. The technique used for the measurement is therefore the following:~1) measure the intensity of $\KPPP$ decay in the $\ks+\kl$ beam as a function of proper $\ks$ time, ~2) measure the same  intensity for a pure $\kl$ beam, ~3) correct the two intensities  for small  differences between beams and systematic effects, ~4) calculate the ratio of intensities and fit the interference term.
\label{data}
\section{Data sample}
\begin{figure}
\resizebox{0.45\textwidth}{!}{%
\includegraphics{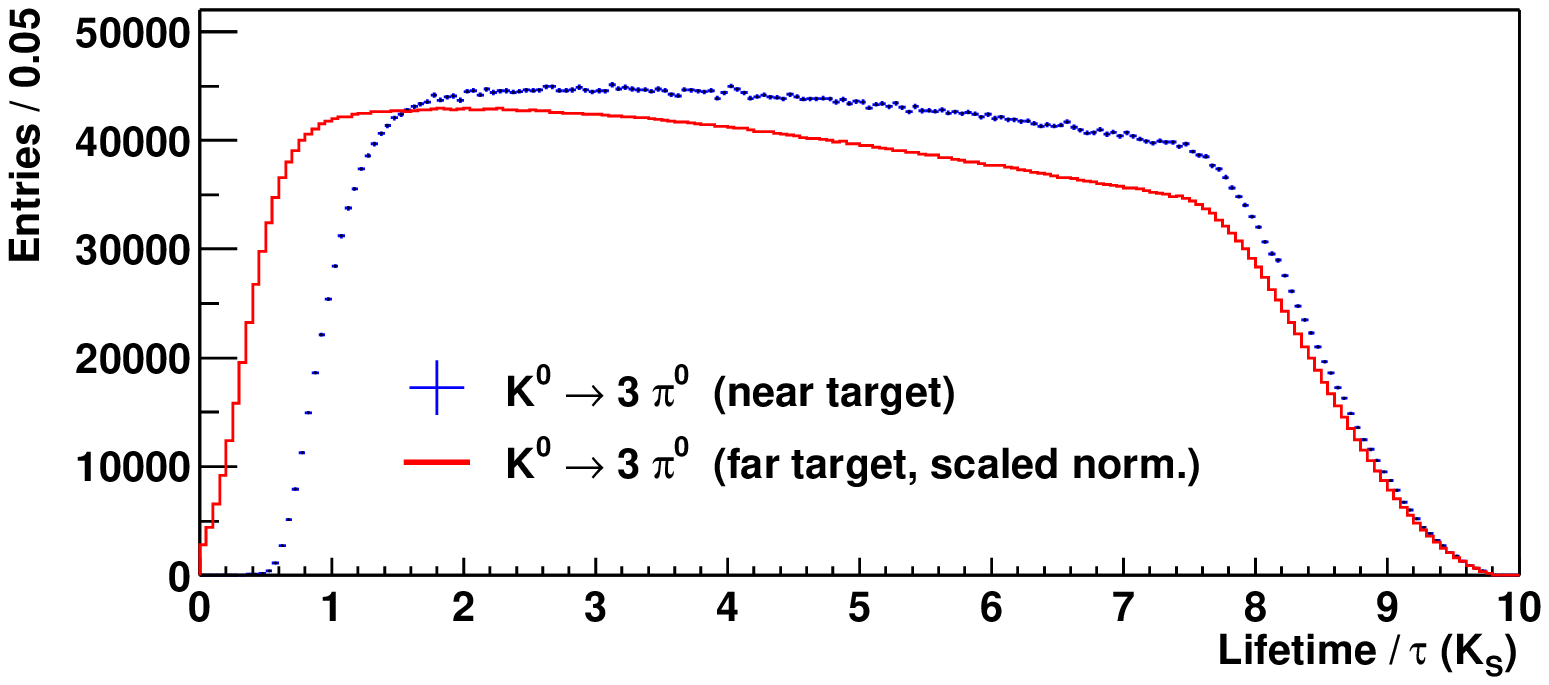}
}
\caption{$\KPPP$ decays from the FAR and NEAR target in unit of $\ks$ proper time}
\label{fig:dist}       
\end{figure}

This analysis have been performed using the data taken during the 2000 run. A sample of $ 6 \cdot 10^{6}~ K_S+K_L \to 3 \pi^0 $ decays from the NEAR target and   $ \sim 10^{7}~ \KLPPP $ decays from the FAR target have been collected, Fig.~\ref{fig:dist}. To extract $\ETAooo$ a fit  to the ratio of the NEAR/FAR samples  is performed in kaon energy bins ($75< E_K < 150$ GeV).  Tab.~\ref{tab:sys} shows the sources of systematic errors. The systematics are dominated by uncertainties in the detector acceptance, accidental activity and the $\kz,\kzb$ dilution.

\begin{center}
\begin{table}
\caption{Source of systematic errors}
\label{tab:sys}
\begin{tabular}{lll} 
 & Re $\ETAooo (10^{-2})$    & Im $\ETAooo (10^{-2})$ \\ 
\noalign{\smallskip}\hline\noalign{\smallskip}
 Accidentals & $\pm 0.1$ &  $\pm 0.6$ \\
 Energy scale & $\pm 0.1$ &  $\pm 0.1$ \\
 Dilution & $\pm 0.3$ &  $\pm 0.4$ \\
 Acceptance & $\pm 0.3$ &  $\pm 0.8$ \\
 Binning & $\pm 0.1$ &  $\pm 0.2$ \\  \hline\hline
 Total & $\pm 0.5$ &  $\pm 1.1$ \\ 
\noalign{\smallskip}\hline
\end{tabular}
\end{table}
\end{center}

\section{Results and discussion}
\begin{figure}[ht]
\resizebox{0.45\textwidth}{!}{%
\includegraphics{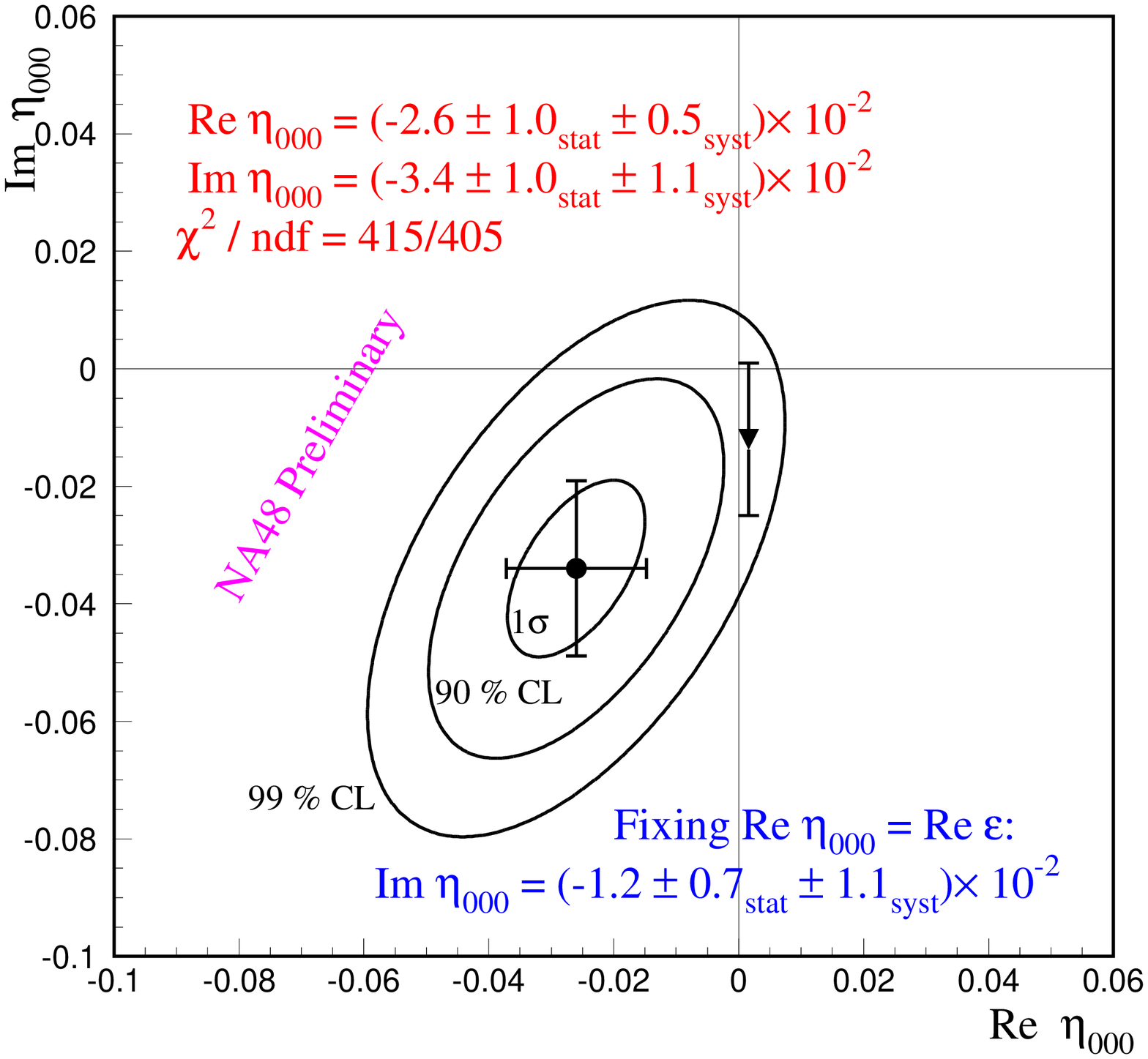}
}
\caption{Fit results for $\ETAooo$ assuming or not $Re(\epsilon) =Re(\ETAooo)$}
\label{fig:res}       
\end{figure}

The result of the simultaneous fit to all energy bins is:
$$Re(\ETAooo) = -0.026 \pm 0.01_{stat}$$
$$Im(\ETAooo) = -0.034 \pm 0.01_{stat}$$
$$Br(\KSPPP) < 1.4 \cdot 10^{-6} \; 90\% CL. $$
The values of $Re(\ETAooo)$ and $Im(\ETAooo)$ have a correlation coefficient of 0.8. According to eq.~\ref{eq:eta} the constrain $Re(\epsilon) =Re(\ETAooo)$ can be used in the fit changing the results to:
$$Im(\ETAooo) = -0.012 \pm 0.007_{stat}\pm 0.011_{sys}$$
$$Br(\KSPPP) < 3.0 \cdot 10^{-7} \; 90\% CL.$$
Fig.~\ref{fig:res} shows these numbers while  Tab.~\ref{tab:oth} lists the results of other experiments. NA48 has improved the precision of both $\ETAooo$ and $Br(\KSPPP)$ by an order of magnitude.

\subsection{The  Bell-Steinberger relation}
Consider a kaon state, superposition of $\ks$ and $\kl$, \\ $ |K(t)> = a_S\ks + a_L\kl$.  Conservation of probability requires that the time derivative of this state is equal to the sum of the decay rates~\cite{th}: $$ -\frac{d}{dt}|<K(0)|K(0)>|^2  =\sum|a_sA(\ks \to f) + a_LA(\kl \to f)|^2.$$
This relation can be rewritten as:
$$ (1+ i~ tan(\phi_{SW}))[Re(\epsilon) - i~ Im(\Delta)] = \sum \alpha_f$$ with $tan(\phi_{SW}) = 2\Delta m/(\Gamma_S -\Gamma_L)$,  $ \alpha_f =  1/\Gamma_S A^*(\ks \to f) A(\kl \to f)$  the possible decays ($\KLPP, ~\KSPPP...$) and $\Delta$ the magnitude of CP violation with CPT violation.   This identity therefore constrains CPT via  the value of $Im(\Delta)$ which, with the new value of $\alpha_{ooo} = \frac{\tau_s}{\tau_L} \ETAooo Br(\KLPPP)$,  is reduced by almost 40\%:
$$Im \Delta = (2.4 \pm 5.0) \cdot 10^{-5} \to Im \Delta = (-1.2 \pm 3.0) \cdot 10^{-5}.$$
$Im(\Delta)$ is now limited by the knowledge of $\eta_{+-}$. Assuming CPT this result can be converted into a new limit on the $\kz,\kzb$ mass difference:
$$ m_{\kz} - m_{\kzb} = (-1.7 \pm 4.2) \cdot 10^{-19} GeV/c^2.$$

\begin{center}
\begin{table}
\caption{Results from other experiments}
\label{tab:oth}
{\tiny
\begin{tabular}{llll}
\noalign{\smallskip}\hline
 Exp. & Year &  Technique &  Result \\ \hline  
 FNAL-E621 & 1994 &$\kz-\kzb$  &  Im $\eta_{+-o} = -1.5 \pm 1.7 \pm 2.5 \cdot 10^{-2}$  \\  & & incoherent & \\ 
 CERN & 1998 &$p -\bar{p} \rightarrow K^-\kzb\pi^+ $  &  Re $\eta_{+-o} =-2 \pm 7^{+4}_{-1}  \cdot 10^{-3}$ \\ CPLEAR & & $\rightarrow K^+\kzb\pi^- $ &  Im $\eta_{+-o} =-2 \pm 9^{+2}_{-1}  \cdot 10^{-3}$ \\ 
Barmin & 1983 &Bubble ch.  &  Re $\eta_{ooo} =-8 \pm 18 \cdot 10^{-2}$ \\ et al. & &  &  Im $\eta_{ooo} =-5 \pm 27  \cdot 10^{-2}$ \\ 
 CERN & 1998 &$p -\bar{p} \rightarrow K^-\kzb\pi^+ $  &  Re $\eta_{ooo} =18 \pm 14 \pm 6  \cdot 10^{-2}$ \\ CPLEAR & &$\rightarrow K^+\kzb\pi^- $ &  Im $\eta_{ooo} =15 \pm 20 \pm 3  \cdot 10^{-2}$ \\ 
Novosibirsk & 1999 &Tagged $K_s$  &  $Br(\KSPPP) < 1.4 \cdot 10^{-5}$ \\ SND & & $ee\rightarrow \phi \rightarrow \ks\ks$ &  \\ 
\noalign{\smallskip}\hline
\end{tabular}
}
\end{table}
\end{center}


\end{document}